# SUPERSYMMETRY SEARCHES AT THE TEVATRON

For CDF and DØ collaborations
R. Demina
*Department of Physics and Astronomy, University of Rochester,*
*Rochester, USA, 14627*

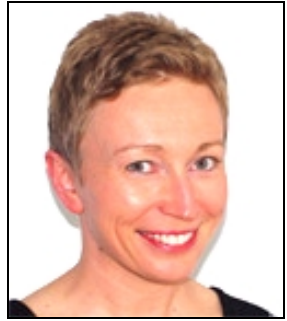

CDF and DØ collaborations analyzed up to 200 $pb^{-1}$ of the delivered data in search for different supersymmetry signatures, so far with negative results. We present results on searches for chargino and neutralino associated production, squarks and gluinos, sbottom quarks, gauge mediated SUSY breaking and long lived heavy particles.

Supersymmetry[1] is a popular extension of the Standard Model originally suggested over 25 years ago. It postulates the symmetry between fermionic and bosonic degrees of freedom. As a result a variety of hypothetical particles is introduced. With presently available experimental data physicists were able to prove that if supersymmetric particles exist they must be heavier than their Standard Model partners[2]. In other words the Supersymmetry is broken. One possible exception is supersymmetric top quark (stop), which still has a chance to be lighter or of the same mass as top quark. With 2-4 $fb^{-1}$ of data Tevatron experiments will be able to extend the limit on stop mass above that of top quark or discover it and thus establish the Supersymmetry[3]. Theory suggests several possible scenarios of Supersymmetry breaking mediated by gravitational or gauge interactions. In gravity mediated scenarios the number of free parameters in the model is reduced to five because of the unification of masses and couplings imposed at the grand unification scale. These parameters are $M_0$ ($M_{1/2}$) – masses of all bosons (fermions) at GUT scale, $A_0$- trilinear coupling and $\mu_0$ – something Higgs and tan($\beta$) – ratio of vacuum expectations of the Higgs doublet.

Up to date CDF and DØ collaborations analyzed up to 200 pb$^{-1}$ of the delivered data in search for different supersymmetry signatures. Weak production of chargino and neutralino (superpatners of W and Z-bosons) can result in a distinct signature – three leptons accompanied by energy misbalance from escaping lightest supersymmetric particles[4]. DØ searched for chargino and neutralino production using the following three signatures: two electrons and an isolated track, electron, muon and an isolated track and two like-sign muons. Search for trilepton signature using two electrons and an isolated track starts with identification of two leptons with the transverse momentum above 12 (8) GeV/c produced in pseudorapidity region of 1.1 (3.0). Invariant mass distribution for these leptons is presented in Figure 1. Contribution from Z-boson decay is reduced by requiring that the invariant mass of the two electrons is between 15 and 60 GeV/c² and that the opening angle between the two electrons is below 2.8 in the transverse plane. Photon production in association with W-boson decaying into an electron and a neutrino can fake dielectron signature if the photon converts into an e⁺e⁻ pair inside the tracker volume. To reduce this possibility DØ required electron track to have at least one hit in the silicon system or to pass tighter electron quality cuts.

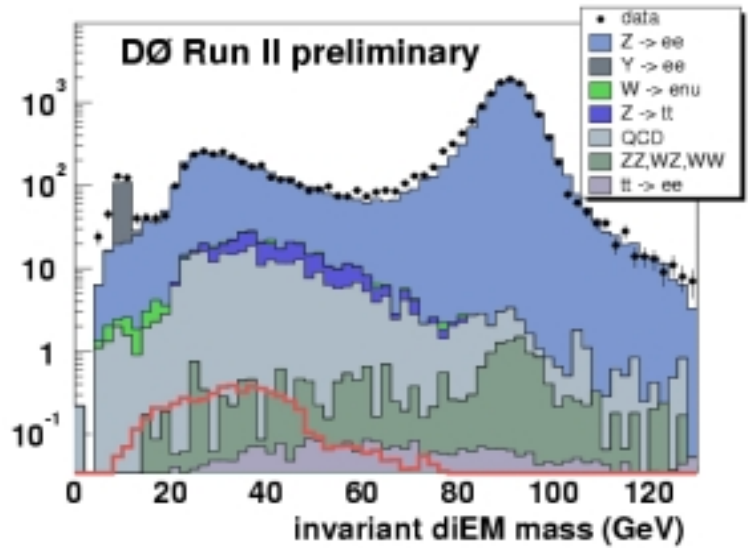

Figure 1. Invariant mass of two electromagnetic objects with the transverse momentum above 12 (8) GeV/c in pseudorapidity region of 1.1 (3.0).

Background from top pair production is reduced by vetoing jets with transverse energy above 80 GeV. The residual background from Drell Yan electron pair production is reduced by requiring a missing transverse energy in access of 20 GeV with an opening azimuth angle of greater than 0.4 with respect to either electron. Third lepton in the event is identified by the presence of an isolated track with transverse momentum greater than 3 GeV/c. Final cut on the cross product of the transverse missing energy and isolated track transverse momentum of above 250 GeV reduces Standard Model background to 0.27±0.42 ±0.02 events. One event is observed in data. Background normalization was verified at several stages of the selection and a good agreement with the data was observed. Similar search strategy was followed in search for the other two signatures of trilepton production. **Table 1** summarizes the number of expected and observed events after the final selection.

Table 1. Number of candidate events observed and background events expected in search for trilepton production.

| Analysis | Data | Total background |
|---|---|---|
| $ee$+isolated track | 1 | 0.27±0.42±0.02 |
| $e\mu$+isolated track | 0 | 0.54±0.24±0.04 |
| $e\mu$ | 1 | 2.49±0.37±0.18 |
| Like sign $\mu\mu$ | 1 | 0.13±0.06±0.02 |

No excess over Standard Model expectation was observed, allowing DØ to set limits on the production of the supersymmetric particles. This limit is presented in Figure 2 as a function of chargino mass. Also presented in this plot is Run 1 DØ result[5]. By using a clever technique of lepton identification as an isolated track DØ was able to extend its sensitivity with respect to the analysis performed on Run 1 data, which searched for three fully identified leptons. Since isolated tracks can be produced by tau-leptons as well as electron and muons, this change in the analyses enhanced DØ sensitivity to tau-enriched signatures which commonly arise in high tan(beta) scenarios. Yet current cross section limit is above theory prediction even for the most optimistic scenario with enhanced leptonic branching fraction due to low slepton masses. Thus no chargino mass range could be excluded by this analysis.

In contrast to weakly produced charginos and neutralinos strongly produced superpartners of quarks and gluons (squarks and gluinos) have large production cross section. The signature of this process is a combination of energetic jets and missing energy[6]. This signature suffers from substantial instrumental backgrounds and stringent cuts must be placed to ensue that background is dominated by physics processes. The analysis starts with a sample dominated by copious two jet production with transverse energy above 60(50) GeV. Jet energy mismeasurement results in missing energy directed along or opposite to one of the jets. To minimize this contribution DØ required that missing energy forms an azimuthal angle between 30 and 165 degrees with the leading jet. Missing transverse energy must exceed 175 GeV and the scalar sum of the transverse energy of all objects in the event must be above 275 GeV. At this point the backgrounds are dominated by vector boson production. 4 events are observed with 2.67±0.95 expected. In $M_0=25$, $M_{1/2}=100$ GeV supersymmetric scenario 17.1 event are expected. In the absence of such excess DØ proceeds with setting limits on the production cross sections of supersymmetric particles, which are presented in Figure 3. Even with 85 pb$^{-1}$ DØ experiment was able to extend its sensitivity beyond that of Run 1[7]. This is largely due to increased beam energy, which leads to a significant increase of the production cross section for heavy objects.

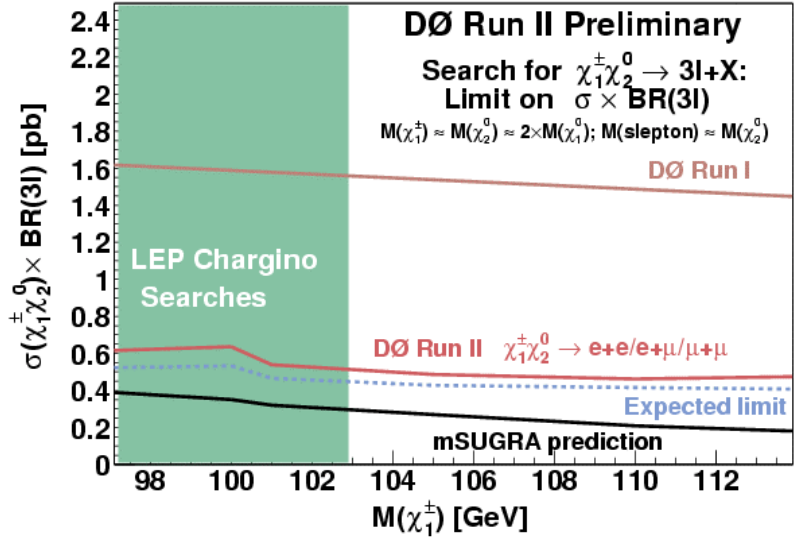

Figure 2. Limits on the total cross section for associated chargino and neutralino production with leptonic final states set by DØ in Run I (top line) and in the current analysis (second from top) in comparison with the expected limit (second from bottom) and the signal cross section predicted for low slepton masses in mSUGRA (bottom line). Chargino masses below 103 GeV are excluded by direct searches at LEP.

CDF combined jets and missing energy signature with heavy flavor tagging to search for gluino cascades to bottom and sbottom quarks[8]. 38 pb$^{-1}$ of data is used to pre-select a sample with three or more jets above 10 GeV and missing transverse energy above 35 GeV. After requiring one or more lifetime based b-jet tags 4 events are observed with 5.6±1.4 expected from the Standard Model sources. One of these events has two tagged b-jets, with the Standard Model expectation of 0.5±0.1 events. As a result a region of sbottom and gluino mass parameter space is excluded, which is presented in Figure 4. With only a fraction of Run 1 integrated luminosity CDF has extended its sensitivity beyond Run 1[9] primarily because of the increased efficiency and acceptance of the silicon system.

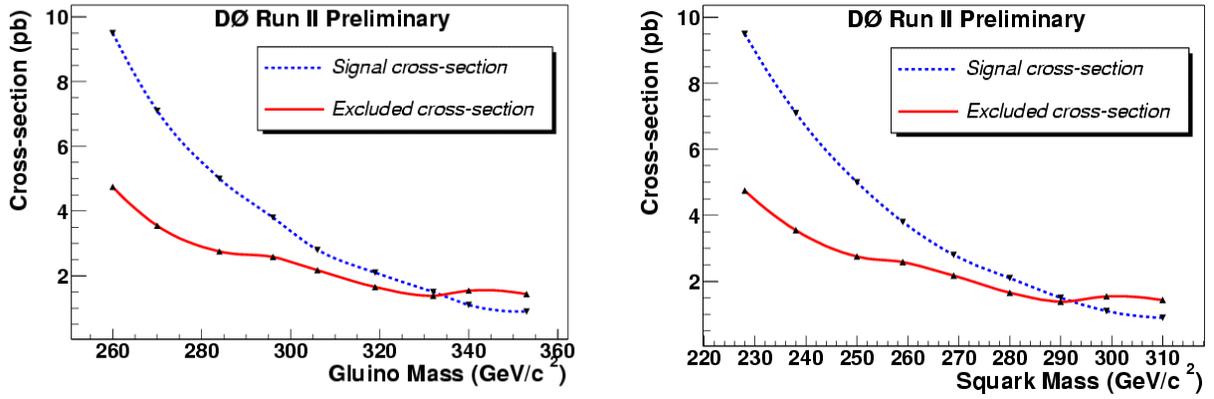

Figure 3. Limits on the production cross sections of gluion (left) and squark (right).

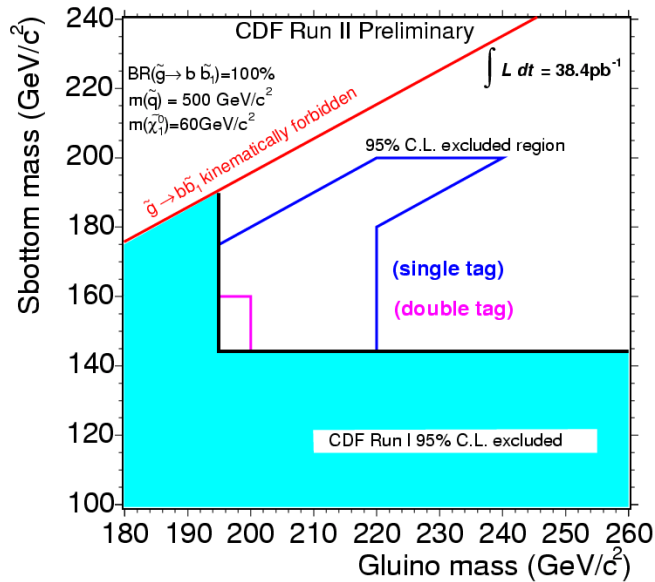

Figure 4. Region of sbottom and gluion parameter space excluded by search for heavy flavor jets produced in association with missing energy.

Gauge mediated SUSY breaking scenario[10] can result in signatures enriched in photons. A sample containing two photons with transverse energy above 20 GeV in pseudorapidity region of 1.1 and missing transverse energy above 40 GeV is searched for signatures of gauge mediated supersymmetry breaking scenarios. 1 event is observed with 2.5±0.5 expected from Standard Model sources, which allowed DØ to limit the scale of gauge mediated SUSY breaking to above 78.8 TeV.

Finally, CDF used its newly acquired time-of-flight system to search for long lived heavy particles. In 53 pb$^{-1}$ of data 2.9±0.7(stat)±3.1(sys) events were expected to traveled longer than 2.5ns from the collision point to the time-of-flight counters, located 1.5m away. 7 events were observed. Because of the large systematic uncertainty on the number of expected events the excess is not considered significant and the result is interpreted as a limit on scalar top production cross section. Two distinctly different supersymmetric scenarios are possible. In one of them scalar top is a single isolated track, in the other it is embedded in the jet of particles produced by scalar top fragmentation. Busy environment in the second scenario reduces the efficiency of scalar top detection and thus the limits on the cross section are less stringent, which is presented in Figure 5. So far, supersymmetry searches yield negative results, but with more data to come these analyses will be updated and new channels will be added to fully exploit Tevatron chances for discovery.

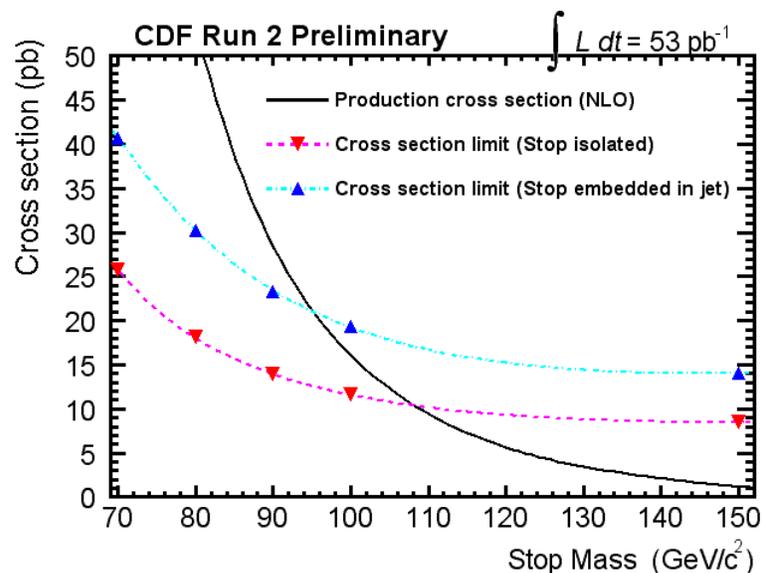

Figure 5. Limit on production cross section of supersymmetric top quark for two scenarios – isolated track and trck embedded in a jet.

We thank the staffs at Fermilab and collaborating institutions, and acknowledge support from the Department of Energy and National Science Foundation (USA), Commissariata l'Energie Atomique and CNRS/Institut National de Physique Nucleaire et de Physique des Particules (France), Ministry of Education and Science, Agency for Atomic Energy and RF President Grants Program (Russia), CAPES, CNPq, FAPERJ, FAPESP and FUNDUNESP (Brazil), Departments of Atomic Energy and Science and Technology (India), Colciencias (Colombia), CONACyT (Mexico), KRF (Korea), CONICET and UBACyT (Argentina), The Foundation for Fundamental Research on Matter (The Netherlands), PPARC (United Kingdom), Ministry of Education (Czech Republic), Natural Sciences and Engineering Research Council and WestGrid Project (Canada), BMBF (Germany), A.P.Sloan Foundation, Civilian Research and Development Foundation, Research Corporation, Texas Advanced Research Program, and the Alexander von Humboldt Foundation.